\documentclass[sigplan]{acmart}

\usepackage{epsfig}
\usepackage{xcolor,color,soul,graphicx}
\graphicspath{ {./figures/} }
\usepackage{comment}

\usepackage{booktabs} % For formal tables
\usepackage{adjustbox}

\usepackage{xcolor,color,soul,graphicx}
\graphicspath{ {./figures/} }

\usepackage[raggedright]{titlesec}
\usepackage{paralist}

\begin{document}

\acmYear{2019}\copyrightyear{2019}
\setcopyright{acmcopyright}
\acmConference[Middleware '19]{Middleware '19: International Middleware Conference Industrial Track}{December 9--13, 2019}{Davis, CA, USA}
\acmBooktitle{Middleware '19: International Middleware Conference Industrial Track, December 9--13, 2019, Davis, CA, USA}
\acmPrice{15.00}
\acmDOI{10.1145/3366626.3368129}
\acmISBN{978-1-4503-7041-7/19/12}

\title[Enabling Enterprise Blockchain Interoperability]{Enabling Enterprise Blockchain Interoperability with Trusted Data Transfer (industry track)}

\author{Ermyas Abebe, Dushyant Behl, Chander Govindarajan, Yining Hu, Dileban Karunamoorthy, Petr Novotny, Vinayaka Pandit, Venkatraman Ramakrishna, Christian Vecchiola}
\affiliation{%
  \institution{IBM Research}
}
\email{etabebe@au1.ibm.com, dushyantbehl@in.ibm.com, chandergovind@in.ibm.com, Yining.Hu@ibm.com, dilebank@au1.ibm.com, P.Novotny@ibm.com, pvinayak@in.ibm.com, vramakr2@in.ibm.com, christian.vecchiola@au1.ibm.com}

\renewcommand{\shortauthors}{IBM Research}

\begin{abstract}

  The adoption of permissioned blockchain networks in enterprise
  settings has seen an increase in growth over the past few
  years. While encouraging, this is leading to the emergence of new
  data, asset and process silos limiting the potential value these
  networks bring to the broader ecosystem. Mechanisms for enabling
  network interoperability help preserve the benefits of independent
  sovereign networks, while allowing for the transfer or sharing of
  data, assets and processes across network boundaries. However, a
  naive approach to interoperability based on traditional
  point-to-point integration is insufficient for preserving the
  underlying trust decentralized networks provide. In this paper, we
  lay the foundation for an approach to interoperability based on a
  communication protocol that derives trust from the underlying
  network consensus protocol. We present an architecture and a set of
  building blocks that can be adapted for use in a range of network
  implementations and demonstrate a proof-of-concept for trusted
  data-sharing between two independent trade finance and supply-chain
  networks, each running on Hyperledger Fabric. We show how existing
  blockchain deployments can be adapted for interoperation and discuss
  the security and extensibility of our architecture and mechanisms.

\end{abstract}




\maketitle

\titlespacing{\section}{0pt}{1.8ex}{0.4ex}
\titlespacing{\subsection}{0pt}{1.5ex}{0.5ex}

\section{Introduction and Motivation}\label{SEC:introduction}
Over the past decade, we have witnessed significant innovation in distributed ledger technologies (DLTs). 
Blockchain technology, a form of DLT, allows a network of independent peer nodes to maintain a consistent view of shared state through consensus, thus enabling decentralized trust. This technology is the basis for well-known decentralized permissionless networks such as Bitcoin~\cite{nakamoto_bitcoin} and Ethereum~\cite{ethereum}. However, applications addressing enterprise use cases impose additional requirements such as scalability, privacy, confidentiality, and auditability. These requirements have led to the development of \textit{permissioned} blockchain networks. Today, there is a wide spectrum of protocol and platform choices for building permissioned networks~\cite{HLF18,CordaTech,quorum,sawtooth}.

Creating blockchain networks entails building the right ecosystem of participants to realize the benefits of decentralization for an application. While these ecosystems should be as broad as possible, the industry trend has been to create networks as {\em minimum viable ecosystems}, i.e., the minimum set of participants required to demonstrate short-term benefits. This trend in conjunction with the wide variety of available platforms has resulted in a landscape of niche, heterogeneous blockchain networks with various methods of data and consensus management, which in turn have led to data and asset silos~\cite{InteropDL}.

To understand the nature and challenge of this fragmentation, let us consider the networks that have emerged in space of global supply chains. We have separate networks for different aspects of the supply chain, like provenance~\cite{ibm_food_trust}, international trade logistics~\cite{tradelens}, trade finance~\cite{wetrade}, and \textit{know-your-customer} (KYC)~\cite{sharedkyc}. An exported product, financed on a trade finance network, is tracked on both a provenance network (for provenance of the product's lifecycle) and a trade logistics network (for the product's shipping status). Meanwhile, the identities of business entities are attested on a shared KYC network. The trustworthiness of a business process on one network can be improved by visibility of data from another network, e.g., the process of fulfilling a \textit{letter of credit} (L/C)~\cite{TFLC} on a trade finance network can be strengthened by fetching the corresponding \textit{bill of lading} (B/L)~\cite{TFBL} from a trade logistics network. However, this is not possible today because data and assets are trapped in their respective networks. The following examples illustrate the varieties of fragmentation. We.Trade~\cite{wetrade} and Marco Polo~\cite{marcopolo}, distinct trade finance networks, exhibit \textit{vertical fragmentation}, being built on different technologies, i.e., Hyperledger Fabric~\cite{HLF18} and Corda~\cite{CordaTech} respectively. We.Trade and TradeLens, an international trade logistics network~\cite{tradelens}, both built on Hyperledger Fabric, exhibit \textit{horizontal fragmentation}, i.e., they support overlapping business processes. Thus, there is a critical need for diverse blockchain networks to interoperate with each other in order to realize greater economic value and potential.

Therefore in this paper, we consider the problem of enabling trusted data exchange between two distinct blockchain networks. We rule out the need to trust individual members in the networks for relaying data, using approaches such as web APIs that act as proxies for individual nodes, since they introduce centralization and consequently pose risk.
A solution can be deemed adequate only if the data exchanged between the networks is accompanied by a proof that represents the consensus view of the network, instead of any individual. 
Such an approach to interoperability has been partially achieved by \textit{relay chains} in Cosmos~\cite{CosmosIBCprotocol} and Polkadot~\cite{DOT19}, into which independent networks can "plug in" and exchange data. However, since neither Cosmos nor Polkadot is universally accepted, a dependency on such frameworks cannot eliminate fragmentation.
Therefore, we present a solution which takes fragmentation into self-sovereign independent networks as a given and build a generic data exchange mechanism into the underlying platforms in such a way that applications running on the platforms can employ these mechanisms to exchange data with another network as per their needs. 

To summarize, we present a network-neutral architecture and implementation for blockchain interoperability. Section~\ref{SEC:modeling} defines interoperability and its challenges. Section~\ref{SEC:architecture} describes a generic system architecture and protocol for trusted data transfer between blockchain networks undergirded by relays and proofs, requiring no trusted mediators. In Section~\ref{SEC:implementation}, we show a proof-of-concept for interoperability based on a use case linking simplified versions of TradeLens and We.Trade. In Section~\ref{SEC:evaluation}, we analyze the security properties of our interoperability protocol, demonstrate the ease of adapting existing applications and show how our architecture is extensible, providing the basis for a universal standard. We discuss related and future work in Sections~\ref{SEC:related-work} and \ref{SEC:conclusion} respectively.

\setlength{\belowcaptionskip}{-15pt}

\section{Interoperability: Definition and Challenges}\label{SEC:modeling}

Abstractly, a blockchain network consists of a set of \textit{peers} that maintain one or more \textit{shared ledgers} (transaction records) without a centralized intermediary. Clients query the ledger and modify it by invoking {\em transactions} that are essentially executions of registered pieces of code called {\em smart contracts}. These transactions are approved and ordered by a consensus protocol into a cryptographically linked chain of \emph{blocks} distributed across multiple peers, thereby ensuring immutability of the ledger data. Permissioned networks built using DLT frameworks like Hyperledger Fabric and Corda also have {\em membership management} modules. Such networks may logically group together clients and peers belonging to the same business entity into {\em organizations}, and may also have \textit{governance frameworks} for managing ledger access and update policies. A blockchain application is typically built on a multi-tier design with smart contracts at the bottom, processing requests from full-stack service applications that run outside the peer network.

The data on the shared ledger may semantically represent state information of the application as well as ownership details of the underlying assets. 
The problem of exchanging assets between different networks has been well studied in the literature and is briefly summarized in Section \ref{SEC:related-work}. However, a generalized protocol for data transfer across DLT implementations along with an ability to provide proofs of the consensus view (i.e., the common view of a representative subset of non-faulty peers) of the transferred data is a relatively unaddressed problem. We define interoperability as \textit{the semantic dependence between distinct ledgers for the purpose of transferring or exchanging data or value, with assurances of validity or verifiability}. We tackle the design of an interoperability framework that supports data transfer across networks in such a way that the consensual validity of the received data can be verified without requiring a centralized intermediary. Henceforth, we will assume that a network has a single ledger, and use the terms "network" and "ledger" interchangeably. In the context of data transfer, we refer to the network sending the data as {\em source} network and the network receiving the data as {\em destination} network.
\begin{figure}
  \centering
  \includegraphics[width=\linewidth]{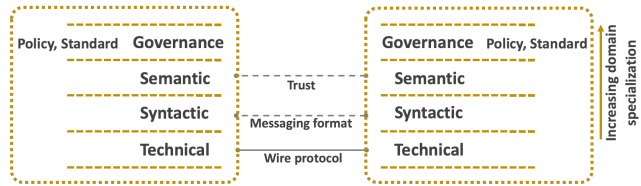}
  \caption{Layered Interaction Model for Blockchain Applications}
  \label{fig:app-layers}
\end{figure}

As illustrated in Figure~\ref{fig:app-layers}, interoperability must be addressed at many levels. However, interoperability challenges for the technical, syntactic, and governance layers in blockchain networks are no different from challenges faced in the Internet and Semantic Web domains. The unique challenges of blockchain interoperability arise at the semantic layer since data is recorded on the shared ledger through consensus among peers maintaining that ledger. Hence, when the destination network receives data from the source network, it needs a mechanism to verify that the data is indeed the consensus view of the source network. Therefore, techniques developed for interoperability at the semantic layer in the Internet and Semantic Web domains where the data source is a centralized system are not applicable to blockchain interoperability.

A framework for enabling interoperation should allow application business logic to be written independent of the framework's implementation. The DLT protocols must therefore offer and use standard abstractions for processes and data while performing certain core functions: \textit{data exposure control}, i.e., making a consensual access control decision in a source network, and \textit{data acceptance} or consensual vetting of data obtained from the source network for authenticity in a destination network. It should be possible to control data exposure and data acceptance policies as per the governance requirements of a network. In general, networks should expose the following operations for interoperability: (i) \textit{query} the data on a source ledger, (ii) carry out \textit{transactions}, or update a source ledger, and (iii) publish and subscribe to \textit{events}. Blockchain applications can exercise these primitives with their business logic.

The practical implementation of a solution to interoperability involves a number of challenges, including specification of a reference architecture and technology-agnostic protocols, avoiding significant modifications to existing applications, discovery of existing networks, and sharing identities and configurations across networks. In this paper, we will tackle the problem of data transfer through cross-network query protocols and define an initial reference architecture for interoperability. In addition, we will demonstrate how existing autonomous networks can be augmented for interoperability with minimal effort. Other aspects, namely cross-network transactions and events, discovery, identity management, and governance, are left to future work.

\titlespacing{\section}{0pt}{2.2ex}{0.5ex}
\setlength{\belowcaptionskip}{-10pt}

\begin{figure*}
  \centering
  \includegraphics[width=\textwidth]{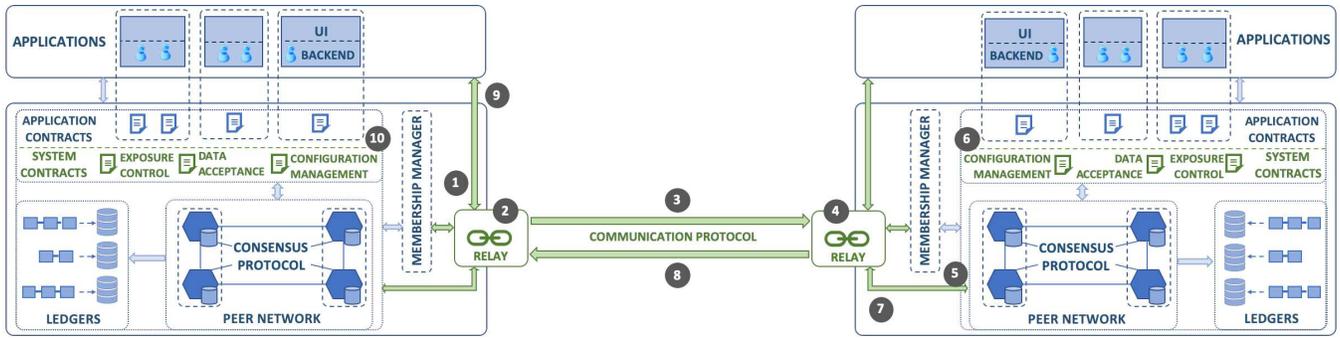}
  \caption{Architecture and Message Flow, with Source Network at Right and Destination at Left (Interoperation Components in Green)}
  \label{fig:arch-interop}
\end{figure*}

\section{Solution For Trusted Data Transfer}\label{SEC:architecture}

We propose an architecture for interoperability between permissioned blockchain networks that extends the generic model sketched in Section~\ref{SEC:modeling} to support cross-network data transfers.

\subsection{Design Principles}\label{arch-prot}

The proposed architecture for trusted data transfer adheres to the following design principles:

\begin{compactitem}
\item Networks are independent self-governing systems that enable interoperability when required.
\item Networks have full control over exposure of data to, and acceptance of data from, other networks.
\item Data transferred must carry verifiable proofs.
\item The network requesting data must be able to specify a policy for proofs (termed as \textit{Verification Policy}) that the source network will satisfy if possible.
\item The communication protocol must be specified in a network-neutral language.
\item Enabling interoperation must not require changes to existing network protocols.
\end{compactitem}

\subsection{Network Components}\label{relay}

Our architecture and message flow for interoperability between two permissioned networks is presented in Figure~\ref{fig:arch-interop}, with the source at the right and the destination at the left. Deployed within, and acting on behalf of, each network is a \textit{\textbf{relay service}} (not to be confused with relay chains in Polkadot~\cite{DOT19} or Cosmos~\cite{ATOM19}). The relay service serves requests for authentic data from applications by fetching the data along with verifiable proofs from remote networks. While the functions of a relay can be implemented on a network's peers, introducing a separate component within the network addresses the design principle of requiring no modifications to network protocols or peer implementations. Relays can thus be deployed and configured within existing networks with ease, and their functionality and communication protocol can evolve independently from any network implementation.

The relays communicate using a shared network-neutral protocol specified using \textit{Protocol Buffers}\cite{protobuf} which enables efficient wire communication. This protocol is structured to provide details on addressing a network, ledger and contract, the function name and arguments for remote queries, a verification policy that is satisfied by the relay in a source network, and authentication details of the requesting entity. Similarly, a response includes the data queried along with a proof that satisfies the verification policy. The relay also includes a set of pluggable \textit{network drivers} that translates the network-neutral protocol messages into calls to the underlying network implementation. The relay thus operates at the technical, syntactic, and semantic layers of the model in Figure \ref{fig:app-layers}.

A set of special \textit{\textbf{system contracts}}, independent of application business logic and deployed on all the peers of the interoperating networks, enforces network rules for data exposure and acceptance. Decisions on what data can be exposed and what acceptance criteria must be applied locally are made by the governing bodies of the respective networks. More specifically, the \textit{Configuration Management} contract maintains identity and configuration (topology) information about foreign networks and is used by other system contracts for every cross-network interaction. The \textit{Exposure Control} contract enforces access control policy rules against incoming requests, determining which data items in the local ledger and smart contract functions can be exposed. The \textit{Data Acceptance} contract enables networks to determine whether the data received from a remote network, along with accompanying proof, is valid according to the verification policy, before it can be written to the local ledger. These system contracts can be implemented and deployed in the same way as application contracts, and thus must be re-implemented for every blockchain implementation using a supported smart contract runtime environment and language.

The architecture assumes minimal trust in the relay. A network can deploy additional relay services to minimize potential censorship attacks and the system contracts can ensure that sensitive data shared between networks is encrypted and unreadable by relays. Furthermore, we assume that interoperating networks have a priori knowledge of each others' identities and configurations, recorded on their ledgers.

\subsection{Message Flow}\label{message-flow}
To enable interoperability between two networks, their system contracts must be initialized with metadata that is determined by the networks' governing bodies and subsequently applied to the respective ledgers by satisfying the networks' consensus rules. In a source network, the \textit{Exposure Control} contract is used to set access control policies at desired granularities. The identities against which the access control policies are applied can be at the level of a network, a named subdivision (organization), a single entity (peer, user or application), or entities that satisfy one or more verifiable credentials. In a destination network, the \textit{Configuration Management} contract defines the verification policies that dictate the criteria under which data and accompanying proof from a remote network must be evaluated for validity by the \textit{Data Acceptance} contract. This verification policy is based on the source network's consensus rules and network units' (e.g., peers) identities; determination of which identities to include in the policy is the responsibility of the destination network.

The steps involved in executing a transaction in a destination network that is dependent on a source network for input data proceed as shown in Figure~\ref{fig:arch-interop}. (1) The client application in the destination submits a request to its local relay service by specifying the source network's unique name, ledger, contract and function to invoke, along with any arguments. It also specifies the verification policy that was determined during the initialization phase. (2) The local relay, designed to support pluggable discovery services, performs a lookup using such a service for the address of the destination relay based on the remote network's name. (3) The destination relay serializes the request and forwards the binary message to the source relay (4) The source relay deserializes the message and determines the network the message is intended for. (5) It then uses the appropriate network driver to orchestrate the query against the respective peers in the network based on the specified verification policy. (6) Each peer executing the contract function refers to the Exposure Control contract to determine if the remote client application has appropriate permissions to read the data. (7) The results from each of the selected peers collectively form the proof satisfying the verification policy. (8) The source relay serializes the results and sends a reply message to the destination relay. (9) The latter then forwards the result back to the application. (10) The application finally constructs a transaction which includes the result of the remote query along with the proof in its input. This transaction is submitted to the appropriate application contract in the local network, which refers to the Data Acceptance contract to validate the query result data and associated proof against the verification policy. If the policy is satisfied, the application contract successfully executes an update to the local ledger.

\section{Use Case and Implementation}\label{SEC:implementation}
The use case we consider is motivated by two industry-scale blockchain consortium networks, TradeLens (TL)~\cite{tradelens} and We.Trade (WT) \cite{wetrade}, both built on Hyperledger Fabric. For simplicity, we constructed scaled-down versions of the two networks using \textit{kubernetes}~\cite{kubernetes}, calling them \textit{Simplified TradeLens} (STL) and \textit{Simplified We.Trade} (SWT) respectively (see Table~\ref{tab:glossary}). While TL connects sellers, carriers, ports, terminal operators and freight forwarders, STL retains just a \textit{Seller} and a \textit{Carrier} negotiating the export of a shipment. WT helps banks facilitate trade financing using \textit{open accounts}~\cite{TFOA}; SWT connects banks and their clients too, but using \textit{letters of credit} (L/C)~\cite{TFLC} (to demonstrate interoperation with STL, as we will see later). Initially built to be non-interoperable, both STL and SWT were augmented to support cross-network queries.

\subsection{Hyperledger Fabric Overview}\label{hlf}

Hyperledger Fabric~\cite{hyperledger_fabric_docs} is an open-source permissioned blockchain platform for building enterprise networks, following the model outlined in Section \ref{SEC:modeling}. Fabric is atypical in its use of an \textit{execute-order-validate} transaction processing model for better privacy and scalability. Though every peer node maintains shared ledger replicas and commits transactions, only a subset run smart contract code (\textit{chaincode}) as \textit{endorsers}. Every transaction requires a subset of endorsers, selected through a predetermined policy, to agree on the result, and a separate \textit{ordering service} creates and disseminates blocks. Peers are classified into \textit{organizations} that represent business interests within a consortium, each typically having its own \textit{Membership Service Provider} (MSP) for identity management and certificate authorization~\cite{HLF18}.

\subsection{Use Case: Supply-Chain and Trade Finance}\label{usecase}

The STL network on Fabric consists of 2 peers: one belongs to a Seller organization and the other to a Carrier organization. A single chaincode manages shipment state and documentation. Independent applications were developed for the Seller and Carrier, invoking chaincode below and offering web UIs above. The SWT network consists of 4 peers: 2 in a Buyer's Bank organization and 2 in a Seller's Bank organization; a Buyer and a Seller are clients of their respective banks' organizations. A single chaincode manages letters of credits and payments. Independent applications were developed for Seller and Buyer.

\begingroup
\setlength{\abovecaptionskip}{5pt}

\begin{figure}
 \includegraphics[width=0.48\textwidth]{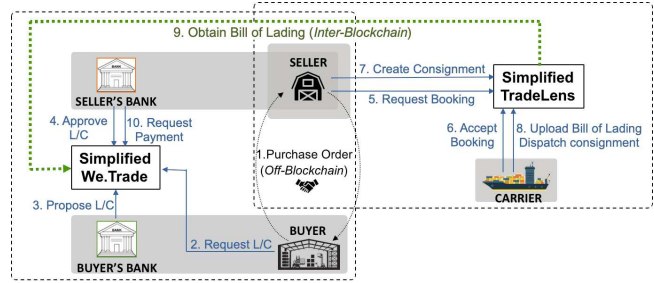}
 \caption{Simplified Trade Interoperation Use Case}
 \label{fig:trade_flow}
\end{figure}

The communication between STL and SWT is illustrated in Figure \ref{fig:trade_flow}. On STL, a seller and a carrier arrange shipment of exported goods against a \textit{purchase order} negotiated offline between the seller and a buyer (Step 1). Steps 5-8 culminate in the carrier taking possession of the shipment and producing a \textit{bill of lading} (B/L)~\cite{TFBL} as proof. On SWT, the buyer's bank issues an L/C for the buyer's transaction against the same purchase order in favour of the seller's bank (Steps 2-4), whose client (seller) is also a member of STL. L/C terms mandate payment upon dispatch; hence the seller's bank may request payment from the buyer's bank as illustrated in Step 10, but it must have proof of existence of a valid B/L, and such proof is fetched from STL using a cross-network query (Step 9.) This interoperation step lets SWT avoid dependence on the seller, who has incentive to forge a B/L and claim payment.
\endgroup

\begin{table}[th]
  \caption{Common Use Case Acronyms}
  \vspace{-0.3cm}
  \begin{adjustbox}{width=0.48\textwidth}
  \centering
  \begin{tabular}{|l|l|}
  \hline
  \textbf{Acronym} & \textbf{Expansion \& Description} \\
  \hline
  L/C & Letter of Credit: Trade Financing Instrument \\
  \hline
  B/L & Bill of Lading: Carrier Acknowledgement of Shipment Receipt \\
  \hline
  (S)TL & (Simplified) TradeLens: Trade Logistics Network \\
  \hline
  (S)WT & (Simplified) We.Trade: Trade Finance Network \\
  \hline
  SWT-SC & Simplified We.Trade-Seller Client \\
  \hline
  ECC & Exposure Control Chaincode \\
  \hline
  CMDAC & Configuration Management \& Data Acceptance Chaincode \\
  \hline
  \end{tabular}
 \end{adjustbox}
 \label{tab:glossary}
\end{table}

\vspace{-0.5cm}
\subsection{End-to-End Data Transfer Protocol}\label{e2e}

\setlength{\abovecaptionskip}{5pt}
\setlength{\belowcaptionskip}{-10pt}
\begin{figure*}[ht]
  \center
  \includegraphics[width=0.95\textwidth]{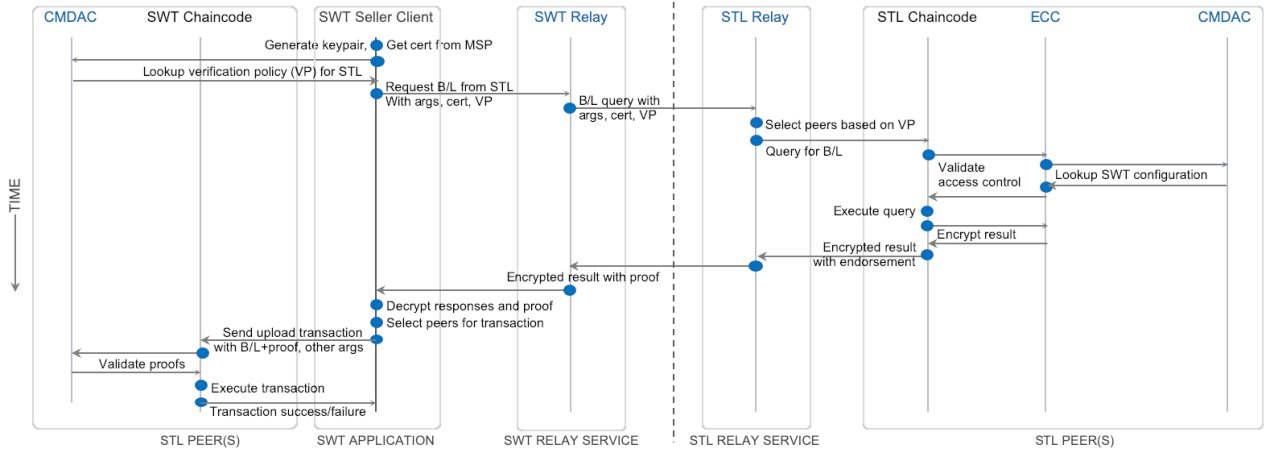}
  \centering
    \caption{Transaction Depending on Cross-Network Query}
    \label{fig:protocol}
\end{figure*}

We implemented a generic protocol for data transfer based on the message flow in Figure~\ref{fig:arch-interop}. Figure~\ref{fig:protocol} illustrates the instance of the protocol used to perform Step 9 in our use case. Before running the applications, both networks were configured in preparation to run cross-network operations. The Exposure Control contract is implemented as an application chaincode (called ECC). The Configuration Management and Data Acceptance contracts are combined into a single application chaincode (called CMDAC) for runtime efficiency, as proof verification depends on foreign networks' configurations. STL configuration (organization and peer identities and root certificates used by MSPs to issue membership credentials) is recorded on the SWT ledger using its CMDAC, as is SWT configuration on the STL ledger. For network discovery and lookup, a local file-based registry was plugged into the SWT Relay.

The STL Chaincode offers a \textit{GetBillOfLading} function that takes a \textit{purchase order reference} as an argument and retrieves the B/L from the ledger. Correspondingly, the SWT Chaincode offers an \textit{UploadDispatchDocs} function that takes a B/L as well as a purchase order reference as arguments. For interoperation, the SWT Chaincode was modified to verify proofs by invoking the CMDAC chaincode. Accordingly, in the SWT Seller application, a \textit{GetBillofLading} query to the relay service was added, along with code to process responses from the relay, decrypt the proof and add the resulting data to the \textit{UploadDispatchDocs} transaction's arguments' list. STL Chaincode was also modified to check if an incoming query is from a relay, and if so, verify access permissions and encrypt the response, all using ECC invocations.

The ECC maintains access control policy rules in the form of a <\textit{network ID}, \textit{organization ID}, \textit{chaincode name}, \textit{chaincode function}> tuple: the subject (requestor) is a member of a network organization and the object (to which the requestor is to be granted access) is a chaincode function. The rule <"\textit{we-trade}", "\textit{seller-org}", "\textit{TradeLensCC}", "\textit{GetBillOfLading}"> was recorded on the STL ledger, which permits STL peers to satisfy B/L queries from members of the Seller's organization in SWT. The CMDAC on the SWT ledger was used to record a verification policy for STL: it requires proof from a peer in both the Seller and Carrier organizations. The (\textit{UploadDispatchDocs}) transaction in SWT requires 2 endorsements: one from a peer each in the Buyer's Bank and Seller's Bank organizations.

We can now follow the data transfer procedure in Figure \ref{fig:protocol}. The SWT Seller Client (SWT-SC) sends a query via its relay to the STL Relay, which in turn orchestrates proof collection by selecting a set of STL peers to query based on the verification policy it receives. Query response (B/L document) and collected proof are returned to the SWT Seller Client via the SWT Relay, which replaces the B/L argument in its transaction request with the received response and proof, and then selects peers to endorse this transaction. (Note that the operations and flows in the Peer(s) sections of Figure \ref{fig:protocol} at both ends occur concurrently on multiple peers.)

For end-to-end confidentiality, the SWT-SC generates an asymmetric key pair and gets a certificate from the Seller organization's MSP. On an STL peer handling the query, the ECC validates the SWT-SC's certificate using the recorded SWT configuration (managed by the CMDAC) before checking whether access ought to be granted as per recorded policy. Post-execution, proof is generated in a 2-step process. First, the result is encrypted with the SWT-SC's public key (from the certificate) using an ECC invocation. Next, the normal peer endorsement process, which produces a signature over query result metadata, is replaced with custom logic~\cite{customescc} that signs the metadata (including the result) and then encrypts it with the SWT-SC's public key. The STL Relay receives an $<$$encrypted\ result$, $encrypted\ metadata$, $signature$$>$ from each STL peer; an array [$<$$encrypted\ metadata$, $signature$$>$] thus constitutes the proof returned to SWT in addition to the $encrypted\ result$. (The result is encrypted to prevent tampering by a malicious relay, whereas the metadata is encrypted to prevent a verifiable proof associated with the result from being exfiltrated by a malicious relay to unauthorized networks; only the SWT-SC possesses a decryption key.) On every SWT peer validating this (decrypted) proof, the chaincode invokes the local CMDAC to validate proof against the verification policy; i.e., validate each signature and authenticate each signer using the recorded STL configuration. Note that replay attacks can be mitigated by \textit{nonces} generated by the SWT-SC and recorded on the destination ledger in every protocol instance.

\titlespacing{\section}{0pt}{1.7ex}{0.5ex}

\section{Evaluation and Discussion}\label{SEC:evaluation}
We have demonstrated how a cross-network data transfer protocol can be designed using system contracts to enforce consensual access control and verification-policy-based proof generation and validation. We now evaluate our design based on the following metrics: (i) security, (ii) ease of use and adaptation, and (iii) generalization and extensibility.

\textbf{Security:}
We evaluate our protocol against the standard information security CIA triad model~\cite{Whitman2011} and reason how it provides \textit{confidentiality}, \textit{integrity} and \textit{availability}. \textbf{(i)} Our protocol satisfies the \textit{\textbf{confidentiality}} property, as response data from source to destination is encrypted by the source's peers using the remote client's public key, ensuring that an untrusted relay cannot read or exfiltrate the information. \textbf{(ii)} It also satisfies the \textit{\textbf{integrity}} property as the source network's peers digitally sign the response and the destination's peers can validate the signatures as well as authenticate the signers using the source network's MSP certificates that are recorded on the destination network's ledger.
\textbf{(iii)} In terms of \textit{\textbf{availability}}, our protocol is not immune to \textit{DoS Attacks}, given the reliance on the relay services of both networks. The effects of DoS attacks can be mitigated by adding redundant relays. DoS protection can also be built into the relay service, protecting the peers themselves from such attacks.

\textbf{Ease of Use and Adaptation:}
The only modification required in a source network is in chaincode. We inserted two function calls in the STL Chaincode: (i) an access control check by invoking the ECC before query execution, and (ii) an encryption call to the ECC after query execution, adding $\sim$35 source lines of code (SLOC). This is a one-time effort, permitting access to functions other than \textit{GetBillOfLading} only requires the addition of a policy rule, and no further chaincode modification.

A destination network requires a more complex configuration, as modifications are needed in chaincode as well as the application.
In the SWT Chaincode, we added $\sim$20 SLOC for proof unmarshaling, and proof verification by invoking the CMDAC; this is a one-time effort. In the higher-layer application, we (i) inserted a remote query call using the relay service API before an \textit{UploadDispatchDocs} transaction submission to the SWT Chaincode and (ii) added calls to decrypt and validate the response and metadata, and run the transaction using the decrypted data and proof as arguments. This required adding $\sim$80 SLOC.
Only a couple of function calls with different parameters need to be inserted in appropriate places in the business logic for other cross-network queries.

\textbf{Generalization and Extensibility:}
The query protocol presented in this paper can be easily extended to enable cross-network chaincode invocations. While the sequence of steps is expected to be different, the relay service, system contracts, and application client support described earlier can be reused directly.

To extend our protocol to other permissioned blockchains, the relay service (which is separate from peer networks) can be directly reused in networks built on Corda or Quorum~\cite{quorum}, for example, with suitable modifications for API compatibility. The system contracts need platform-specific implementations.
In Corda, a verification policy can be specified to include signatures from \textit{notaries}~\cite{notaries_corda}, which will be involved in access control, proof generation and verification. In Quorum, proof generation may require augmenting a peer to return a signed query response in addition to implementing our system contracts. 
The functions served by these contracts will remain the same, as will the relay communication API offered by the target platform's SDK. These networks have identity managers and certificate authorities as Fabric does, so network configuration can be shared for proof validation and authentication.
Standardization of platform-independent schemas will be required to facilitate the sharing of these configurations.

\section{Related and Complementary Work} \label{SEC:related-work}

\textbf{\indent Cross-Network Communication Protocols:}
Many emerging projects are actively developing inter-blockchain protocols to build \textit{networks of networks}. The most prominent among them, Cosmos~\cite{ATOM19} and Polkadot~\cite{DOT19}, enable cross-network transactions between separate blockchains, called \emph{zones} in Cosmos and \emph{parachains} in Polkadot, via a central blockchain, called the \emph{hub} in Cosmos and \emph{relay chain} in Polkadot. Cosmos and Polkadot networks are decentralized: i.e., every transaction committed to the central blockchain is verified by a group of validators.
Polkadot requires all participating networks to conform to a common consensus protocol to achieve inter-network operations. Cosmos allows participating blockchains to "plug in" to the main hub keeping their original consensus protocol. However, Cosmos requires all participating networks to comply with the Inter-Blockchain Protocol (IBC)~\cite{CosmosIBCprotocol} for interoperability.
Polkadot and Cosmos are also not interoperable with each other, forcing existing network pairs to choose and implement the same protocol for mutual interoperability.
In comparison, our design does not impose any such constraints on participating networks or reliance on a central blockchain.

\textbf{Cross-Network Data Transfer with Proof:}
A number of mechanisms to prove the veracity of transactions recorded on a ledger using self-sufficient proofs have been proposed for both public~\cite{peace_relay,kiayias2016PoPoW,kiayias2017NiPoWPoW} and private~\cite{ATOM19,DOT19,CordaBIP} blockchains, creating a basis for trusted data transfer.
Cosmos, Polkadot and Corda all utilize mechanisms in which a subset of members attest their approvals of a transaction, thereby proving its validity. SPV-based~\cite{peace_relay, zamyatinxclaim, kiayias2018proof}, PoPoW~\cite{kiayias2016PoPoW} and NIPoPoW~\cite{kiayias2017NiPoWPoW} proof mechanisms are used to prove the validity of a transaction in a public ledger (e.g., Bitcoin, Ethereum) on any other blockchain by certifying the work done for that particular transaction (in other words: \textit{proving proof-of-work}).
Our protocol implementation uses a form of attestation-based proof, but the architecture allows any suitable proof scheme to be plugged in. Furthermore, our architecture enables consensual data exposure control and acceptance, which none of the aforementioned proof schemes provide on their own.

\textbf{Cross-Network Asset Transfers:}
Atomic asset transfers and exchanges between ledgers are an important interoperation scenario for which techniques like \textit{atomic swaps}~\cite{herlihy2018atomic} and \textit{Hash Time Locked Contracts (HTLC)}~\cite{HTLC} have been invented. In addition, networks implementing the Interledger Protocol (ILP)~\cite{ILPv4} can swap assets across multiple blockchain network hops, each hop swapping the asset using an HTLC variant. In the future, we will consider incorporating these techniques into our architecture and protocol to enable a wider spectrum of applications including both asset and data transfers.

\titlespacing{\section}{0pt}{2ex}{1ex}

\section{Conclusion and Future Work}\label{SEC:conclusion}

Enabling seamless interoperability is critical to realizing the potential of blockchain technology. In this paper, we have taken a concrete step towards presenting a reference architecture and protocol for trusted data transfer between blockchain networks using local relays and policy-driven control of interactions. We have also demonstrated proof of concept based on a real industry use case. We plan to implement protocols for cross-network transactions and events and incorporate asset exchanges to enable a wider array of use cases. We will leverage technologies like decentralized identifiers~\cite{DID19} and verifiable credentials~\cite{VCVP19} to build decentralized frameworks for the sharing of network identities and configurations. We will also investigate the construction of an optimal verification policy from a network's consensus policy.

\section*{Acknowledgments}\label{SEC:acknowledgments}

The authors would like to thank Adarsh Saraf and Bruno Marques for helping with system design, Isabell Kiral and Nick Waywood for helping build the applications and Allison Irvin and Isabell Kiral for review and feedback.

\bibliographystyle{ACM-Reference-Format}
\bibliography{middleware-industry}

\end{document}